\documentclass[preprint,12pt]{elsarticle}

\usepackage{amssymb}
\usepackage{amsmath}

\usepackage{color}
\usepackage[colorlinks=true, linkcolor=blue, citecolor=blue, urlcolor=blue]{hyperref}
\usepackage{multirow}

\journal{arxiv}

\tnotetext[ieeenote]{This work has been submitted to the IEEE for possible publication. Copyright may be transferred without notice, after which this version may no longer be accessible.}

\begin{document}

\begin{frontmatter}



\title{
Perturbation-Based Pinning Control Strategy for Enhanced Synchronization in Complex Networks
}

\author[inst1]{Ziang Mao}

\affiliation[inst1]{organization={University of Science and Technology of China},
            addressline={School of Cyber Science and Technology}, 
            city={Hefei},
            postcode={230000}, 
            state={Anhui},
            country={China}}

\author[inst1]{Tianlong Fan\corref{cor1}}
\cortext[cor1]{Corresponding authors}
\ead{tianlong.fan@ustc.edu.cn}

\author[inst1]{Linyuan {Lü}\corref{cor1}}

\ead{linyuan.lv@ustc.edu.cn}

\begin{abstract}

Synchronization is essential for the stability and coordinated operation of complex networked systems. Pinning control, which selectively controls a subset of nodes, provides a scalable solution to enhance network synchronizability. However, existing strategies face key limitations: heuristic centrality-based methods lack a direct connection to synchronization dynamics, while spectral approaches, though effective, are computationally intensive. To address these challenges, we propose a perturbation-based optimized strategy (PBO) that dynamically evaluates each node's spectral impact on the Laplacian matrix, achieving improved synchronizability with significantly reduced computational costs (with complexity $O(kM)$). Extensive experiments demonstrate that the proposed method outperforms traditional strategies in synchronizability, convergence rate, and pinning robustness to node failures. Notably, in all the empirical networks tested and some generated networks, PBO significantly outperforms the brute-force greedy strategy, demonstrating its ability to avoid local optima and adapt to complex connectivity patterns. Our study establishes the theoretical relationship between network synchronizability and convergence rate, offering new insights into efficient synchronization strategies for large-scale complex networks.

\end{abstract}



\end{frontmatter}


\section{Introduction}
\label{sec:introduction}

Synchronization among networked systems is critical to ensuring stability and coordinated behavior in various applications, including distributed computing, multi-agent systems~\cite{chen2015saturated}, power grids~\cite{dorfler2013synchronization}, and information dissemination~\cite{zanette2002dynamics}. In many instances, only a subset of nodes needs to be controlled to induce global synchronization, a concept known as pinning control~\cite{sorrentino2007controllability, yu2013step,yu2013synchronization}. 
Traditional approaches for selecting pinning nodes typically rely on static centrality measures such as degree, betweenness~\cite{brandes2001faster}, or eigenvector centrality~\cite{wang2010control}. However, these metrics often fail to capture the complex structural and dynamical properties inherent in large-scale heterogeneous networks, resulting in suboptimal performance. Moreover, the driver nodes selected by these centrality measures frequently exhibit a pronounced ``rich club'' effect, tending to cluster in the network’s core; as a result, their collective influence in driving global synchronization is diminished. In addition, these heuristic methods~\cite{fan2021characterizing,qiu2021identifying} predominantly consider only the topological features of nodes and lack a direct correlation with the synchronization dynamics—such as the critical role played by the smallest eigenvalue of the Laplacian matrix in determining network synchronizability.

In this work, we address these limitations by introducing a perturbation-based greedy algorithm that leverages matrix perturbation theory~\cite{stewart1990stochastic} to quantify the impact of controlling individual nodes on the network's spectral properties. By iteratively assessing the influence of each candidate node on the smallest eigenvalue of the grounded Laplacian matrix~\cite{miekkala1993graph,pirani2014spectral}, our approach dynamically identifies the nodes that most effectively enhance both network synchronizability and convergence speed. Furthermore, when applied to both synthetic and real-world networks, our approach consistently outperforms traditional methods, demonstrating its effectiveness and robustness across diverse network topologies and real-world complexities.

The remainder of this paper is organized as follows. Sec.~\ref{sec:related} reviews relevant literature on pinning control and node selection strategies. Sec.~\ref{sec:preliminaries} introduces the fundamental concepts of network synchronization and pinning control. In Sec.~\ref{sec:methodology}, we detail the proposed perturbation-based node selection algorithm and its theoretical underpinnings. Sec.~\ref{sec:experiments} details our experimental setup and Sec.~\ref{sec:results} presents comparative results. Sec.~\ref{sec:discussion} describes related discussions of our experiments. Finally, Sec.~\ref{sec:conclusion} concludes the paper and outlines future research directions.

\section{Related Work}
\label{sec:related}
Research on network synchronization and pinning control has steadily evolved over the past decades. Early studies, such as those by Li \emph{et al.}~\cite{li2004pinning}, demonstrated that controlling high-degree nodes can stabilize scale-free networks, while random pinning may suffice for homogeneous networks. Subsequent work by Yu \emph{et al.}~\cite{yu2009pinning} showed that networks can be synchronized under linear feedback schemes by appropriately adjusting the coupling strength. Although these static and adaptive approaches have advanced the field, they are inherently limited by their reliance on fixed topological metrics and the need for precise parameter tuning.

More recent research has shifted toward spectral-based methods that exploit the eigenstructure of the Laplacian matrix to guide pinning node selection. For example, Almendral \emph{et al.}~\cite{almendral2007dynamical} explored the link between network dynamics and spectral properties, while Amani \emph{et al.}~\cite{amani2018controllability} and Jalili \emph{et al.}~\cite{jalili2015optimal} developed strategies based on the augmented Laplacian. Other studies have formulated the pinning control problem using semidefinite programming to derive optimal feedback gains~\cite{jafarizadeh2021optimal,jafarizadeh2022pinning}, and Cheng \emph{et al.}~\cite{cheng2018selecting} introduced a method based on the left Perron vector. Despite their theoretical advantages, these spectral-based approaches often entail high computational costs when applied to large and complex networks.

Additional methodologies have been proposed to handle practical challenges such as time delays, stochastic disturbances, and uncertainties. Intermittent pinning control schemes~\cite{liu2015synchronization,xu2017finite,ding2020synchronization} reduce control effort by applying signals aperiodically, while event-triggered approaches~\cite{adaldo2015event, li2018event,xing2020robust} and fuzzy control strategies~\cite{li2016fuzzy,nguyen2019fuzzy} aim to enhance robustness. However, these dynamic methods typically depend heavily on extensive parameter tuning, which can compromise their reliability in real-world scenarios.

\section{Preliminary}
\label{sec:preliminaries}
In this section, we briefly review the basic concepts of network synchronization and pinning control.

\subsection{Network Synchronization}
Consider an undirected and unweighted network \(G\) composed of \(N\) identical oscillators. The dynamics of the \(i\)-th node is described by
\begin{equation}  \label{eq:sync_dynamics}
\begin{split}
\dot{x}_i &= F(x_i) + c\sum_{j=1}^{N} a_{ij}\left[H(x_j)-H(x_i)\right] \\
&= F(x_i) - c\sum_{j=1}^{N} \ell_{ij} H(x_j) \\
&= F(x_i) - c\sum_{j=1}^{N} \ell_{ij} \Gamma x_j,
\end{split}
\end{equation}
where \(i=1,2,\dots,N\), \(x_i \in \mathbb{R}^n\) denotes the state vector of node \(i\), \(F(\cdot)\) describes the intrinsic dynamics, \(H(\cdot)\) is an output function, \(c>0\) is the coupling strength, \(A=(a_{ij}) \in \mathbb{R}^{N\times N}\) is the adjacency matrix, \(L=(\ell_{ij}) \in \mathbb{R}^{N\times N}\) is the Laplacian matrix (positive semi-definite), and \(\Gamma\) is the inner coupling matrix.

Let the target state \(s(t)\), which may represent an equilibrium point, a periodic solution, or a chaotic orbit, satisfy
\begin{equation}\label{eq:target_state}
\dot{s}(t)=F(s(t)),\quad s(0)=s_0.
\end{equation}
The objective of synchronization is to ensure that
\begin{equation}\label{eq:synchronization_goal}
\lim_{t\to\infty} \|x_i(t)-s(t)\|=0,\quad i=1,2,\dots,N.
\end{equation}

\subsection{Pinning Synchronization}
To achieve network synchronization, only a fraction $\delta(0<\delta<1)$ of nodes (denoted by \(i_1, i_2, \dots, i_l\)) are directly controlled by applying a linear feedback signal. The dynamics of the controlled nodes are given by
\begin{equation}
\dot{x}_{i_k}=F(x_{i_k})-c\sum_{j=1}^{N}\ell_{i_kj}\Gamma x_j+u_{i_k},\quad k=1,2,\dots,l,
\end{equation}
where \(u_{i_k}\) is the feedback control signal. Without loss of generality, assume the first \(l\) nodes are the pinning nodes. The overall controlled system can then be expressed as
\begin{equation}
\dot{x}_i=F(x_i)-c\sum_{j=1}^{N}\ell_{ij}\Gamma x_j-c\,d_i\,\Gamma(x_i-s),\quad i=1,2,\dots,N,
\end{equation}
with the control gain \(d_i\) defined by
\begin{equation}
d_i=\begin{cases}
>0, & i=1,2,\dots,l,\\[1mm]
0, & i=l+1, \dots, N.
\end{cases}
\end{equation}
In our analysis, we assume that the control gains for the pinning nodes are constant and identical.

\subsection{Stability of Pinning Synchronization}
Let \(x_i(t)=s(t)+\delta x_i(t)\) and linearize \(F(x_i)\) around \(s(t)\) so that
\[
F(x_i)=F(s)+DF(s)\,\delta x_i.
\]
Defining the error vector \(\xi=x-s\) and the diagonal control gain matrix \(D=\text{diag}(d_1,d_2,\dots,d_N)\), the linearized error dynamics become
\begin{equation}\label{eq:linearized}
\dot{\xi}=[DF(s)-c(L+D)\Gamma]\xi.
\end{equation}
Projecting \(\xi\) onto the eigenbasis of the augmented Laplacian \(\bar{L}=L+D\) (i.e., \(\bar{L}\phi_i=\lambda_i\phi_i\)) and writing \(\xi=\Phi v\) (with \(\Phi=[\phi_1, \phi_2, \dots, \phi_N]\)), we obtain
\begin{equation}
\dot{v}=[DF(s)-c\Gamma\Lambda]v,
\end{equation}
where \(\Lambda=\text{diag}(\lambda_1,\lambda_2,\dots,\lambda_N)\). Consequently, the evolution of each mode is governed by
\begin{equation}
\dot{\xi}_h=[DF(s)-c\lambda_h\Gamma]\xi_h,\quad h=1,2,\dots,N.
\end{equation}
Using the Master Stability Function (MSF) framework~\cite{pecora1998master}, synchronizability of the coupled system is characterized by the region where \(\lambda_{max}(c\lambda_h)<0\).

In the Type II synchronization regime, the unbounded synchronization region is characterized by the smallest nonzero eigenvalue of the augmented Laplacian matrix \(\bar{L}\)~\cite{arenas2008synchronization}. A necessary condition for stable synchronization is then given by
\begin{equation}\label{eq:type2}
\lambda_{1}(\bar{L}) > \frac{\alpha_1}{c},
\end{equation}
where \(\alpha_1\) is determined by the intrinsic node dynamics and the coupling function. When the control gain is appropriately chosen, it can be shown that \(\lambda_1(\bar{L})\) is equivalent to \(\lambda_1(\hat{L}_{N-l})\)~\cite{liu2018optimizing}, where \(\hat{L}_{N-l}\) denotes the submatrix of \(L\) corresponding to the unpinned nodes; hence, \(\lambda_{1}(\hat{L}_{N-l})\) serves as an indicator of synchronizability of the coupled system.

While Type I synchronization—characterized by a bounded synchronization region \(c\lambda_h \in (\alpha_1,\alpha_2)\)~\cite{pecora1998master,arenas2008synchronization, donetti2005entangled, tang2014synchronization}—is well-documented, it primarily applies to systems with finite control parameter ranges. Given that unbounded synchronization regions (Type II) are prevalent and practically relevant for large-scale networks, our analysis and subsequent algorithm focus exclusively on the Type II regime.

\subsection{Convergence Rate of Pinning Synchronization}\label{subsec:convergence}
To evaluate convergence rate of the system, which measures the speed at which the system evolves from the initial state to the target state $s$, let $W_h=||\xi_h||^2=\xi_h^T\xi_h$ \cite{yan2009synchronization}. Assuming $\Gamma$ is the identity matrix, the rate of change of \(W_h\) is given by
\begin{equation}
\frac{\dot{W}_h}{W_h}=\frac{\xi_h^T\left[(DF(s))^T+DF(s)\right]\xi_h}{\|\xi_h\|^2}-2c\lambda_h.
\end{equation}
If the fluctuation term is relatively small, then
\begin{equation}
W_h(t)\approx W_h(0)e^{-2c\lambda_ht},
\end{equation}
and the overall convergence rate is determined by \(\min_{h}2c\lambda_h=2c\lambda_1\).

Thus, the smallest eigenvalue of the augmented Laplacian, \(\bar{L}\), governs the convergence rate of the pinning synchronization process. When the control gains are appropriately chosen such that \(\lambda_1(\bar{L}) = \lambda_1(\hat{L}_{N-l})\), an increase in \(\lambda_1(\hat{L}_{N-l})\) not only improves synchronizability but also accelerates the convergence rate.

\section{Methodology}
\label{sec:methodology}

We propose an efficient node selection strategy for pinning control based on spectral properties and matrix perturbation theory. In addition to a traditional brute-force greedy approach (used as an optimal baseline), our method employs a perturbation index to significantly reduce computational complexity.

\subsection{Brute-Force Greedy Strategy}\label{subsec:brute}
The brute-force greedy strategy iteratively selects the node that maximizes the increase in the smallest eigenvalue of the grounded Laplacian matrix \(\hat{L}\). The procedure is as follows:

\begin{enumerate}
    \item Initialization: Let the network \(G(V,E)\) be represented by its Laplacian matrix \(L\) and initialize the pinning set \(S=\emptyset\).
    \item Iteration: For each node \(v\in V-S\), compute the smallest eigenvalue \(\lambda_1(\hat{L}_{V-S\cup\{v\})}\)
    of the submatrix obtained by removing rows and columns corresponding to \(S\cup\{v\}\). Select the node \(v^*\) that maximizes the increase:
    \[
    v^* = \arg\max_{v\in V-S} \lambda_{1}(\hat{L}_{V-S\cup\{v\}}).
    \]
    Add \(v^*\) to \(S\).
    \item Termination: Repeat until the number of selected nodes reaches the preset threshold \(k\) (typically 10\%--30\% of \(N\)).
\end{enumerate}

The greedy strategy is effective because it dynamically and adaptively evaluates the impact of enclosing each node on the overall synchronizability of the network. Although this dynamic evaluation leads to a more optimized selection process than heuristic centrality methods, it also incurs a significant computational cost.

\subsection{Optimized Strategy with Perturbation Index}\label{subsec:optimized}

To mitigate the high computational cost inherent in the greedy strategy, we introduce a perturbation index that quantifies the influence of each node on the smallest eigenvalue. We approximate the change in \(\lambda_1\) by applying matrix perturbation theory~\cite{stewart1990stochastic}. 

Specifically, consider a general matrix \(M\) (e.g., the grounded Laplacian matrix) and perturb it by virtually removing the \(i\)th node, which results in a small change \(\delta M\),
\begin{equation}
\delta M = M' - M = -M_{:,i}e_i^T - e_i M_{i,:} + M_{i,i}\,e_i e_i^T,
\end{equation}
where \(M_{:,i}\) and \(M_{i,:}\) denote the \(i\)-th column and row of \(M\), respectively, and \(e_i\) is the \(i\)-th standard basis vector. Using first-order perturbation theory,
\begin{equation}\label{eq:delta_lambda}
\delta \lambda_{1} \approx u^T\delta M\,u = -2\,u_i\,(u^T M_{:,i}) + M_{i,i}\,u_i^2,
\end{equation}
where \(u\) is the unit eigenvector corresponding to \(\lambda_1\) and \(u_i\), its \(i\)-th component, quantifies the contribution of the \(i\)-th node to \(\lambda_1\). Noting that

\[
u^T M_{:,i} = (Mu)_i = \lambda_{1}\,u_i,
\]

we obtain

\begin{equation}
\delta \lambda_{1} \approx u_i^2\left(M_{i,i}-2\lambda_{1}\right)>0.
\end{equation}

For the grounded Laplacian matrix \(\hat{L}\), \(M_{i,i}\) equals the degree \(d_i\) of the \(i\)-th node. We then use \(\delta\lambda_1(i)=u_i^2(d_i-2\lambda_1)\) as a measure of the importance of node \(i\).

The optimized node selection process is as follows:
\begin{enumerate}
    \item Initialization: Represent the network \(G(V,E)\) by its Laplacian matrix \(L\) and initialize \(S=\emptyset\).
    \item Iteration: Compute the smallest eigenvalue \( \lambda_1(\hat{L}_{V-S}) \) of the current submatrix \( \hat{L}_{V-S} \) and its corresponding eigenvector \( u \). Using this information, determine the perturbation index \( \delta \lambda_1(v) \) for each node \( v \in V-S \). Select the node \[
    v^* = \arg\max_{v\in V-S} \delta \lambda_1(v),
    \] and add it to \( S \).
    \item Termination: Stop when \(k\) nodes have been selected.
\end{enumerate}

This strategy allows us to estimate the impact of each node on \(\lambda_1\) quickly by avoiding the full eigenvalue recomputation for every unselected node at every iteration, thereby greatly reducing the computational complexity while still closely approximating the performance of the optimal greedy method.

\section{Experimental Setup}\label{sec:experiments}
We evaluate the proposed perturbation-based strategy against traditional centrality-based methods (degree and betweenness) and the optimal brute-force greedy strategy across various networks. The evaluation focuses on synchronizability, convergence rate, and robustness under node failures.

\subsection{Network Generation}

Three types of synthetic networks are generated:

\begin{enumerate}
    \item Scale-Free Networks (BA): Generated using the Barabási–Albert model~\cite{albert2002statistical}, exhibiting a power-law degree distribution \(P(k)\sim k^{-\gamma}\).
    \item Random Networks (ER): Constructed using the Erdős–Rényi model, where each pair of nodes is connected with probability \(p_1=0.1\).
    \item Small-World Networks (WS): Generated using the Watts–Strogatz model~\cite{watts1998collective} by rewiring edges in a regular ring lattice with probability \(p_2=0.1\).
\end{enumerate}

In addition, five real-world networks spanning biological, social, and human-contact domains~\cite{rossi2015network} are evaluated:
\begin{enumerate}
    \item \textit{C. elegans}: The metabolic network characterizes substrate-level biochemical interactions~\cite{duch2005community}.
    \item DD-g1076: A biological network representing protein-protein interactions.
    \item Twitter-Copen: A social medial network, which captures retweet and mention interactions generated during the United Nations Climate Change Conference in Copenhagen~\cite{ahmed2010time}.
    \item Infect-Dublin: A physical human contact network where nodes correspond to individuals and edges represent direct physical contacts~\cite{infect}.
    \item Socfb-Haverford: A social network derived from Facebook, which represents social friendship ties among individuals~\cite{traud:2011fs,traud2012social}.
\end{enumerate}

The basic statistics of the datasets are shown in Table~\ref{Datasets}.

\begin{table}[htbp]
    \centering
\caption{The basic statistics of the real datasets. \(N\) and \(M\) represent the total number of nodes and edges respectively}
\label{Datasets}
    \begin{tabular}{cccccc} \hline  
        
        Dataset& \(N\)& \(M\)&Max degree & Min Degree&Average Degree\\ \hline  
        
        \textit{C. elegans}& 453&2040 & 237& 1&8\\  
        
        DD-g1076& 891& 1859 & 20& 2&8\\   
        
        Twitter-Copen& 761& 1029 &37 & 1&2\\   
 Infect-Dublin&410&2765 & 50& 1&13\\ 
 Socfb-Haverford& 1446& 59589& 375& 1&82\\ \hline
    \end{tabular}
    
\end{table}

\subsection{Node Pinning Strategies}
We compare the following methods:
\begin{enumerate}
    \item Degree-Based Strategy (Degree): Selecting the \(k\) nodes with the highest degree.
    \item Betweenness-Based Strategy (Betweenness): Selecting the \(k\) nodes with the highest betweenness centrality.
    \item Brute-Force Greedy Strategy (BFG): This optimal baseline strategy is detailed in Sec.~\ref{subsec:brute}.
    \item Perturbation-Based Optimized Strategy (PBO): As described in Sec.~\ref{subsec:optimized}.  
\end{enumerate}
Each strategy iteratively selects and pins nodes until the preset number \(k\) is reached. And we list the time complexity of each methods, as shown in Table~\ref{TimeComplexity}. Specifically, for the BFG strategy, at each of the \( k \) iterations, all remaining candidate nodes (approximately \( O(N) \)) need to be evaluated by recomputing the smallest eigenvalue, and each eigenvalue computation costs approximately \( O(M) \) operations for sparse networks. Therefore, the overall complexity is calculated as $O(kNM)$. In contrast, for the proposed PBO strategy, each iteration involves only one eigenvalue computation and the calculation of perturbation indices for all remaining nodes, with each perturbation index requiring constant-time computation. Thus, the complexity per iteration is dominated by eigenvalue computation as \( O(M) \), resulting in a significantly lower overall complexity $O(kM)$.

\begin{table}[htbp]
    \centering
    \caption{The time complexity of pinning nodes selection strategies. \(k\), \(N\) and \(M\) respectively represent the required number of pinning nodes, the total number of nodes and the total number of edges.}
    \begin{tabular}{cc} 
        \hline  
        Pinning control strategy& Time complexity\\ \hline
          
        Degree& \(O(N)\)\\ 
          
        Betweenness& \(O(NM)\)\\ 
 BFG&\(O(kNM)\)\\  
 PBO&\(O(kM)\)\\ \hline\end{tabular}
    \label{TimeComplexity}
\end{table}

\section{Results and Analysis}
\label{sec:results}
\subsection{Synchronizability}
\begin{figure}[htbp]
    \centering
    \includegraphics[width=\linewidth]{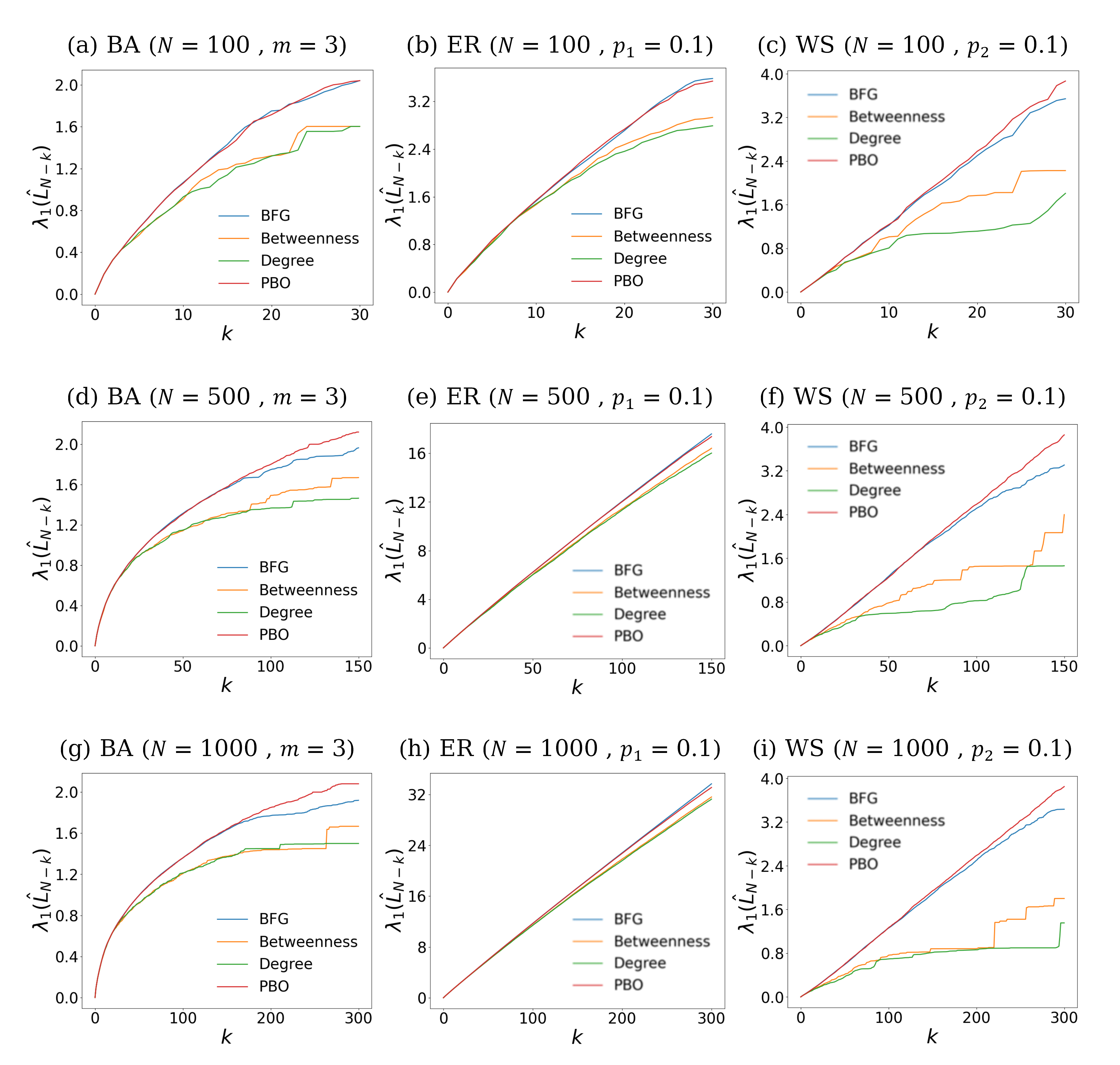}
    \caption{Synchronizability of node pinning strategies in synthetic networks of varying sizes. The variable $k$ represents the number of pinning nodes, while the titles above each subplot indicate the size of the synthetic networks and the generation parameters. Different solid lines correspond to different pinning strategies. }
    \label{fig:synchronizability}
\end{figure}

The synchronizability of a network is evaluated by observing the change in the smallest eigenvalue \(\lambda_1\) of the grounded Laplacian matrix as pinning nodes are incrementally added. Figure~\ref{fig:synchronizability} shows the evolution of \(\lambda_1\) for BA, ER, and WS networks. Our results indicate that, across all network types, \(\lambda_1\) generally increases with the number of pinning nodes, reflecting enhanced control over the network dynamics. 
In particular, the brute-force greedy strategy (BFG), which serves as the benchmark for synchronization performance, consistently yields higher \(\lambda_1\) values than traditional centrality-based methods. Remarkably, the proposed perturbation-based strategy (PBO) not only significantly outperforms traditional centrality-based approaches, but also surpasses the BFG on several networks across different scales, highlighting its strong effectiveness and potential advantages over computationally intensive baseline.

Compared with the three network models, the PBO and BFG strategies exhibit remarkable consistency in ER networks which can be explained by its relatively homogeneous structure.
However, in the BA and WS networks, while PBO approximates BFG well for small values of \(k\), its performance experiences a significant increase compared to BFG as \(k\) increases. The enhanced performance of PBO in BA networks can be attributed to its ability to identify critical nodes in a more optimized way, which compensates for the structural heterogeneity in BA networks, where high-degree hubs dominate the structure, but low-degree nodes actually determine the upper bound of global synchronizability~\cite{liu2018optimizing}. In WS networks, PBO again outperforms BFG, showcasing its effectiveness in networks with small-world properties.
These findings highlight both the advantages of the PBO strategy and the challenges posed by the structural characteristic of various networks.

\subsection{Convergence Rate}
To evaluate the convergence rate of the proposed perturbation-based strategy, we examine the synchronization time (SyncTime) required for the network to evolve from random initial states to the synchronized target state \(s\). A shorter SyncTime indicates a faster convergence rate and consequently better synchronizability.

We employ Chen's system~\cite{zhou2008pinning} to model the individual node dynamics, where the state of each oscillator is governed by:
\[
\begin{bmatrix}
\dot{x}_1\\
\dot{x}_2\\
\dot{x}_3
\end{bmatrix}
=
\begin{bmatrix}
p_1(x_2-x_1)\\
(p_3-p_2)x_2-x_1x_3+p_3x_2\\
x_1x_2-p_2x_3
\end{bmatrix},
\]
with parameters \(p_1=35\), \(p_2=3\), and \(p_3=28\). In these settings, the system exhibits chaotic behavior with an unstable equilibrium point. To stabilize the system to the target state $s$, we apply pinning control to a subset of nodes by introducing a linear feedback signal, and the corresponding controlled dynamics are implemented as described in Sec.~\ref{sec:preliminaries}.

For consistency, key parameters such as the coupling strength \(c\) and control gain \(d_i\) are held constant across simulations for all compared strategies. The synchronization time is computed as the average time over multiple trials.

Table~\ref{synctime} summarizes the average smallest eigenvalue \(\lambda_1\) of the grounded Laplacian and the corresponding average synchronization time for three types of networks with 1000 nodes. The results indicate that the average convergence rate of the PBO is slightly faster than the BFG, and greatly superior to those of the Degree and Betweenness methods across different networks. Moreover, synchronization occurs faster in networks with higher \(\lambda_1\) values, showing that there is a clear consistency between the smallest eigenvalue and the synchronization time. This observation validates our theoretical derivations, confirming that \(\lambda_1\) serves as a reliable proxy for both the synchronizability and convergence rate of the network.

\begin{table}[htbp]
    \centering
    \caption{Comparison of average synchronization time (SyncTime) and smallest eigenvalue $\lambda_1$ across different networks and strategies. The table demonstrates the relationship between $\lambda_1$ and the average SyncTime, where higher $\lambda_1$ values correlate with shorter synchronization times.}
    \begin{tabular}{cccc} 
        \hline  
        Network Type & Strategy & $\lambda_1$ & Average SyncTime \\ 
        \hline  
        \multirow{4}{*}{BA(1000)} & Degree & 1.4498 & 0.4806\\ 
         & Betweenness & 1.4393 & 0.4915\\ 
         & BFG & 1.7714 & 0.3798\\ 
         & PBO & 1.8497& 0.3633\\ 
        \hline  
        \multirow{4}{*}{ER(1000)} & Degree & 21.5818 & 0.0236\\ 
         & Betweenness & 21.8455 & 0.0233\\ 
         & BFG & 22.8160 & 0.0225\\ 
         & PBO & 22.7241& 0.0226\\ 
        \hline  
        \multirow{4}{*}{WS(1000)} & Degree & 0.8615& 0.8889\\ 
         & Betweenness & 0.8842& 0.8057\\ 
         & BFG & 2.4973& 0.2434\\ 
         & PBO & 2.5856& 0.2295\\ 
        \hline 
    \end{tabular}
    \label{synctime}
\end{table}

\subsection{Robustness}
\begin{figure}[htbp]
    \centering
    \includegraphics[width=1\linewidth]{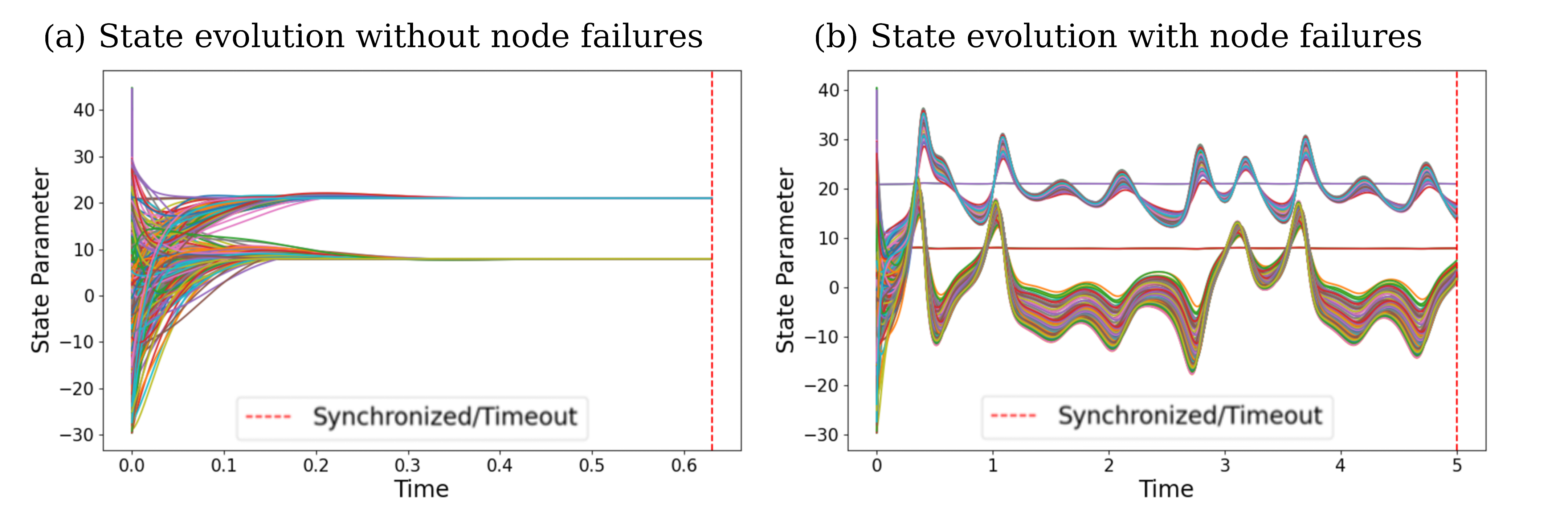}
    \caption{State evolution of individual nodes over time, with each line corresponding to a different node. The red vertical dashed line indicates the synchronization time or timeout point, where nodes either achieve synchronization or fail to synchronize within the given time. In panel (b), node pinning failures are introduced, leading to more variability in node behavior and a delayed synchronization process.}
    \label{fig:trajectory}
\end{figure}

In practical scenarios, pinning nodes may be vulnerable to failures or attacks, which can significantly impair the network's controllability~\cite{li2020secure}. To quantify this effect, we examine the controllability robustness (CR) of our pinning control system, where CR is defined as the sequence of values that record the remaining controllability after a sequence of
attacks~\cite{lou2021knowledge}. Under attack, the dynamics of the controlled system are modified as follows:
\begin{equation}
\dot{x}_i = F(x_i) - c \sum_{j=1}^{N} \ell_{ij} \Gamma x_j - c\,d_i\,(1-\beta_i) \Gamma (x_i-s),
\end{equation}
where \(\beta_i\in\{0,1\}\) indicates whether node \(i\) has failed (with \(\beta_i=1\) meaning that the node is inactive). Figure~\ref{fig:trajectory} illustrates a representative case of the nodal states evolution in a scale-free network (100 nodes) with and without node pinning failures. 

To evaluate robustness, we simulate scenarios in which a fixed fraction of the pinning nodes is randomly deactivated, thereby mimicking potential pinning failures. For each strategy, the smallest eigenvalue \(\lambda_1\) of the grounded Laplacian is computed after removing the corresponding rows and columns of the remaining active pinning nodes. This eigenvalue serves as an indicator of the network’s synchronizability under failure conditions, a higher \(\lambda_1\) implies better robustness. 

\begin{figure}[htbp]
    \centering
    \includegraphics[width=\linewidth]{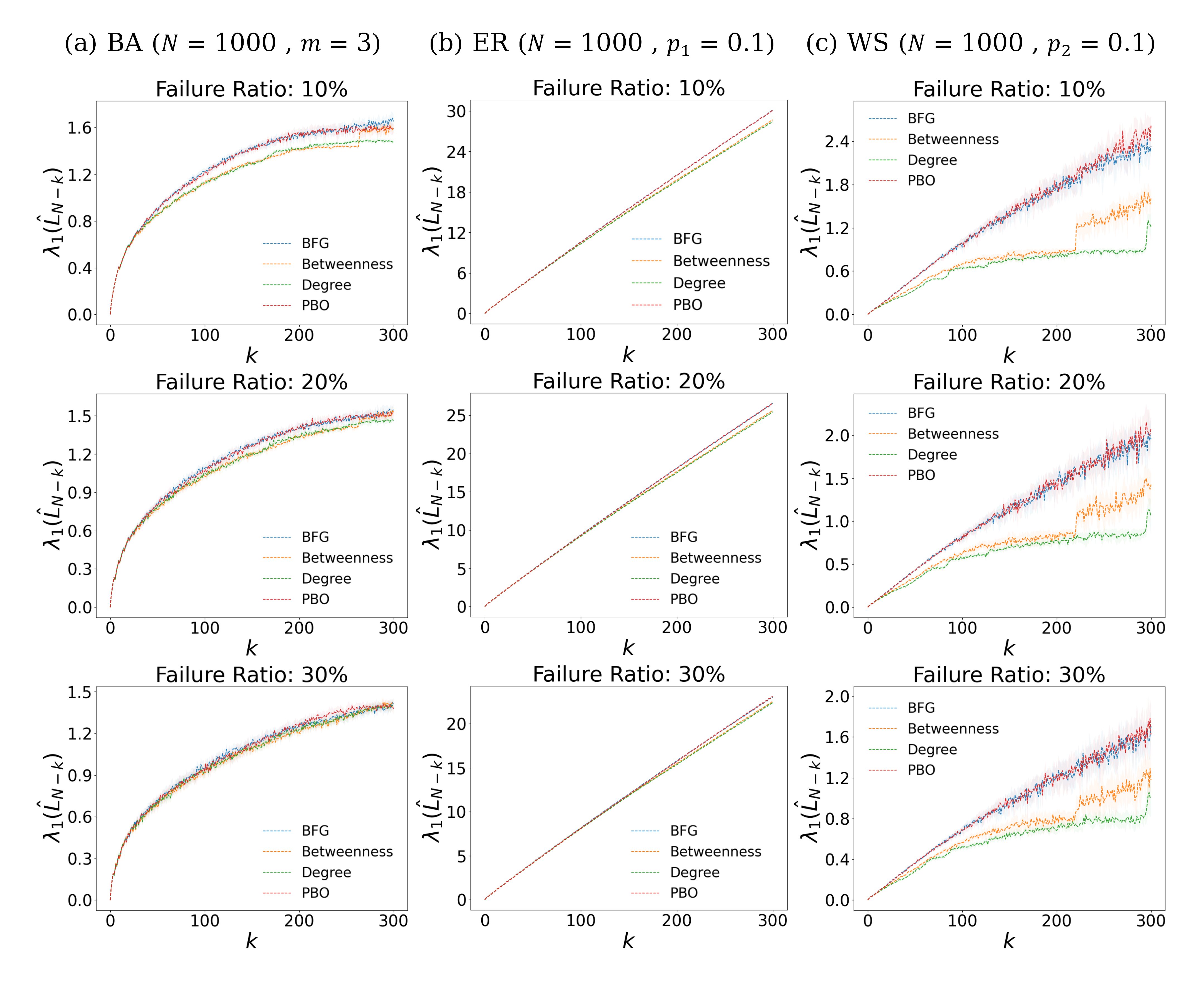}
    \caption{Synchronizability of node pinning strategies under different pinning node failure ratios (10\%, 20\%, 30\%) across BA, ER, and WS networks (1000 nodes). The variable $k$ represents the number of pinning nodes, while the titles above each subplot indicate the proportion of failed pinning nodes relative to the total pinned nodes. Each column of three subplots corresponds to the same network, with the network configuration displayed at the top. Different dashed lines correspond to different pinning strategies, and the shaded regions represent the standard deviation.}
    \label{fig:robustness}
\end{figure}

To mitigate the influence of random variations, the deactivation process is repeated across multiple trials for each failure ratio. Specifically, for each selected pinning node set, the corresponding failure ratio is simulated by randomly deactivating a designated percentage (e.g., 10\%, 20\%, or 30\%) of the pinning nodes. The resulting $\lambda_1$ values are then averaged over these trials to obtain a robust estimate. As shown in Figure~\ref{fig:robustness}, while $\lambda_1$ continues to increase with the number of pinning nodes $k$, the rate of improvement diminishes as the failure ratio increases, in contrast to the results in Figure~\ref{fig:synchronizability}. Notably, the PBO strategy consistently maintains higher $\lambda_1$ values than traditional centrality-based methods, demonstrating superior synchronizability. Moreover, PBO remains well aligned with BFG, and in certain scenarios, it even surpasses BFG, highlighting its robustness in preserving network synchronizability despite pinning node failures.

\subsection{Analysis of Real Networks}

\begin{figure}[htbp]
    \centering
    \includegraphics[width=\linewidth]{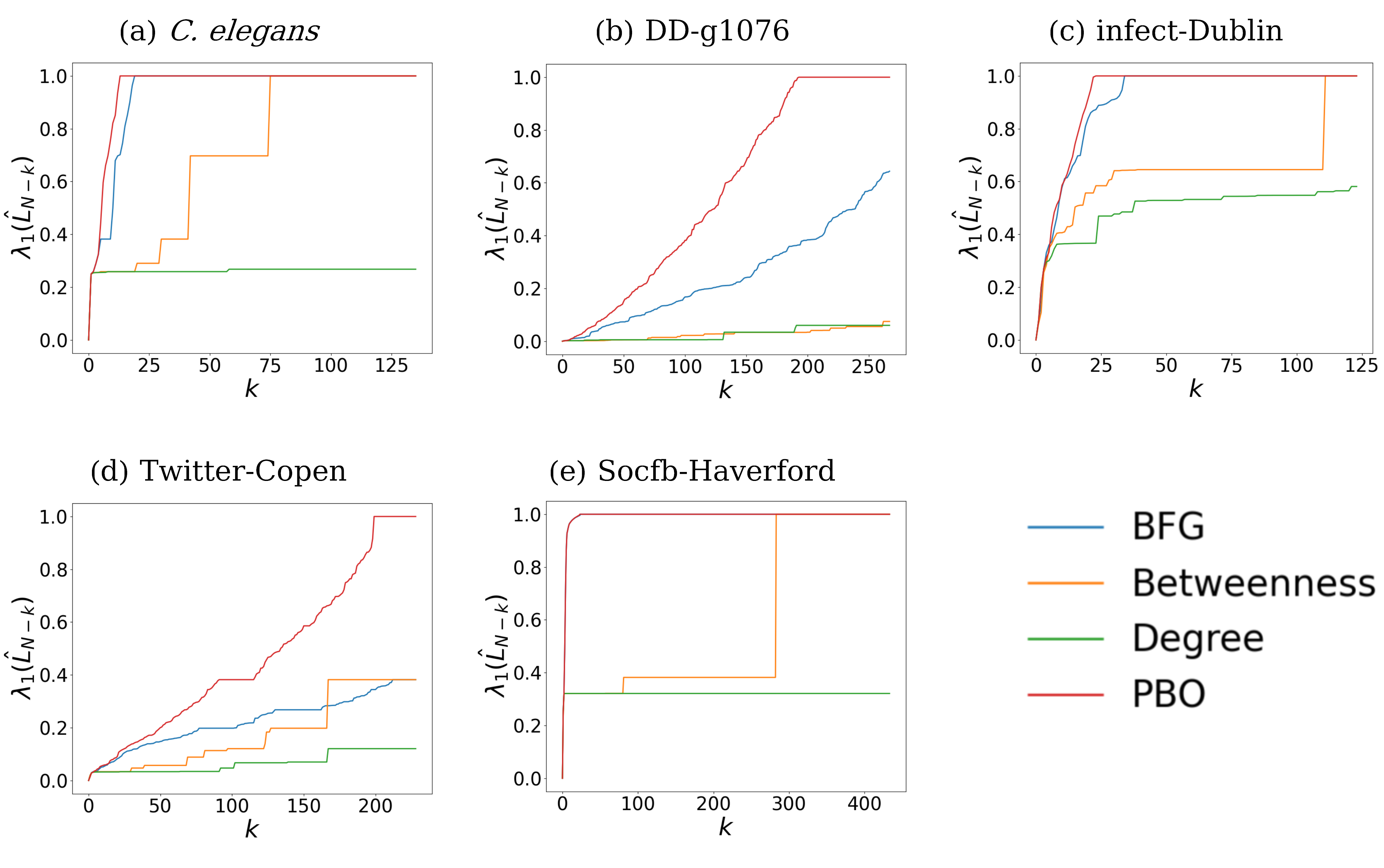}
    \caption{Synchronizability of node pinning strategies across the five real-world networks. The variable $k$ represents the number of pinning nodes, while different solid lines correspond to different pinning strategies.}
    \label{fig:real_network_synchronizability}
\end{figure}

We further validated the proposed strategy by evaluating its pinning effectiveness and robustness across multiple real-world networks spanning biological, social, and human-contact domains~\cite{rossi2015network}. Figure~\ref{fig:real_network_synchronizability} presents the results of synchronizability under pinning control. Notably, PBO consistently outperforms the baseline BFG strategy and others across nearly all real-world networks. This advantage is particularly pronounced in networks such as DD-g1076 and Twitter-Copen, where PBO exhibits a substantial lead in enhancing synchronizability. In the Socfb-Harverford network, PBO well approximates the BFG and achieves high synchronizability way more faster than the centrality-based strategies.

This superiority of PBO over BFG in real-world networks may be attributed to several key factors. First, PBO avoids the limitations of global greedy search, which may lead to suboptimal selections in networks with intricate topologies. Second, PBO considers both the eigenvalue impact and node degree influence through the perturbation index \( \delta \lambda_1(v) \), while also leveraging the leading eigenvector \( u_i \) to reflect each node's importance in the synchronization process, as its magnitude often indicates the node's global influence. Third, the structural heterogeneity of real-world networks, characterized by diverse connectivity patterns and strong community structures, reduces the effectiveness of conventional greedy optimization. 

By leveraging spectral perturbation analysis, PBO dynamically adapts to network-specific characteristics, enabling it to identify critical nodes that exert maximal influence on synchronization. These advantages distinguish PBO from both heuristic centrality-based methods and purely greedy optimization approaches, demonstrating its broader applicability in complex, real-world settings.

It is worth noting  that the \(\lambda_1\) values across all these real-world networks in Figure~\ref{fig:real_network_synchronizability} converge to approximately 1. This phenomena can be attributed to the sparsity of these networks, where many candidate nodes have only one neighbor. As the synchronizability is upper-bounded by the minimum degree among the unpinned nodes~\cite{liu2018optimizing}, such structural sparsity inherently limits the achievable improvement in \(\lambda_1\).

\begin{figure}[htbp]
    \centering
    \includegraphics[width=\linewidth]{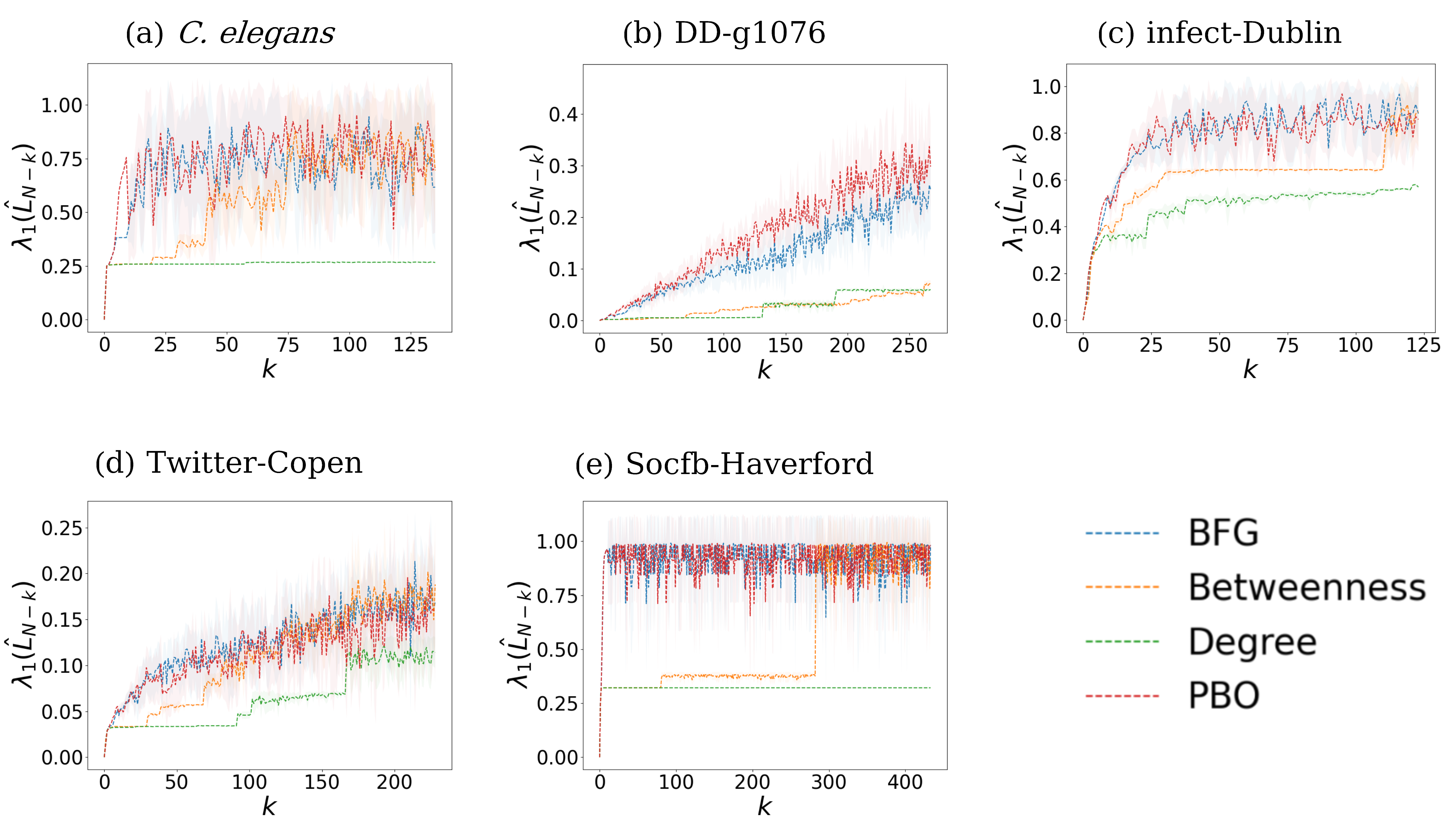}
    \caption{Synchronizability of node pinning strategies under node failure ratio 10\% across real-world networks. The variable $k$ represents the number of pinning nodes, while different dashed lines correspond to different pinning strategies, and the shaded regions represent the standard deviation.}
    \label{fig:real_network_robustness}
\end{figure}

Figure~\ref{fig:real_network_robustness} further illustrates the robustness of pinning strategies under node failures. As expected, synchronizability declines in all cases compared to failure-free conditions. However, PBO and BFG maintain significantly higher synchronizability levels than the Degree and Betweenness strategies, reaffirming their superior effectiveness and robustness even under adverse conditions.

The weak performance of Degree and Betweenness strategies likely stems from their reliance on static topological rankings, which fail to capture the intricate interplay between network structure and synchronization dynamics. As demonstrated in both figures, these heuristic approaches consistently lag behind, reinforcing the necessity of incorporating spectral properties into pinning selection. The substantial gap between PBO and traditional strategies highlights the critical role of optimization-driven node selection in real-world synchronization tasks, particularly in scenarios involving node failures and evolving network dynamics.

\subsection{Impact of Network Sparsity}

\begin{figure}[htbp]
    \centering
    \includegraphics[width=\linewidth]{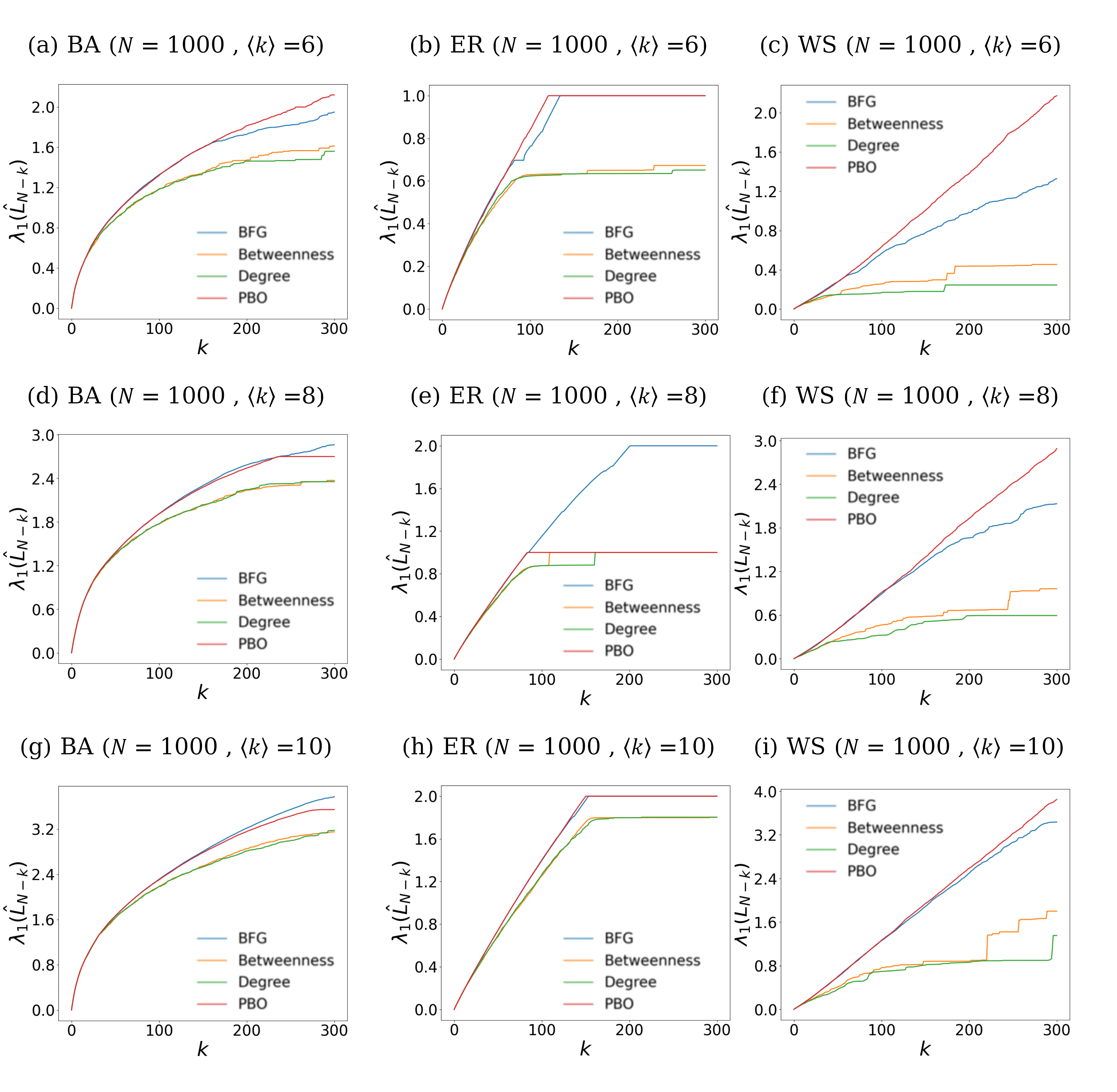}
    \caption{Synchronizability of node pinning strategies across synthetic networks with various densities. The variable $k$ represents the number of pinning nodes, while the titles above each subplot indicate the size and average degree of the synthetic networks. Different solid lines correspond to different pinning strategies.}
    \label{fig:sparsity}
\end{figure}

To investigate the impact of network sparsity on pinning synchronizability, we further evaluate effectiveness of the compared strategies across different networks with various densities, by constructing BA, ER and WS networks with the average degree \(\langle k\rangle=6,8,10\), as shown in Figure~\ref{fig:sparsity}.

We observe that as the network density increases, the performance gap between the proposed PBO strategy and BFG baseline generally narrows across different networks, except for the ER network with \(\langle k\rangle=6\) , probably due to its inherently higher structural randomness. Notably, PBO slightly underperforms BFG on BA network with \(\langle k\rangle=8, 10\). These results suggest that in sparser networks, where node influence varies significantly, PBO consistently outperforms the BFG and centrality-based approaches. However, as density increases, structural redundancy and enhanced connectivity reduce the sensitivity of synchronizability to the selection of specific nodes, thereby diminishing the relative advantage of PBO. Despite this, PBO still outperforms Degree and Betweenness-based methods across all conditions. Overall, these findings underscore the robustness and adaptability of the proposed strategies, particularly highlighting its effectiveness in sparser networks, which dominate real-world social, biological, and communication systems.

\section{Discussion}
\label{sec:discussion}
Our experimental results reveal that the set of pinning nodes selected by the proposed perturbation-based strategy exhibits minimal overlap with those chosen by traditional centrality-based methods, underscoring the method's ability to identify influential nodes that are otherwise overlooked. This observation highlights a key limitation of conventional centrality measures, which tend to rely on local topological features and do not necessarily correlate with the global synchronizability of the network.

Furthermore, our findings show that the synchronizability generally increases with the number of pinning nodes. However, in heterogeneous networks such as those generated by the Barabási–Albert model, \(\lambda_1\) tends to saturate for larger fractions of pinned nodes—likely due to the inherent upper bound imposed by the minimum degree of the unpinned nodes. This suggests that, to further enhance synchronization in such networks, it may be necessary to consider strategies that also incorporate low-degree nodes. 

Notably, The BFG strategy, which was expected to deliver the best performance by exhaustively searching for the optimal pinning nodes, surprisingly underperformed compared to the Perturbation-Based Optimized (PBO) strategy on some generated networks and certain real-world networks. A significant portion of this result can be attributed to the specific topological characteristics of these networks, e.g sparsity and heterogeneity. This observation further highlights the effectiveness and practicality of the PBO strategy, due to its better adaptability to some real networks. However, the exact reasons why PBO outperforms BFG in these cases—specifically how certain network features influence PBO’s success—remain an open question that warrants further exploration.

\section{Conclusion}
\label{sec:conclusion}
In this study, we addressed the challenge of optimizing synchronization in complex networks by proposing a novel perturbation-based greedy algorithm for selecting pinning control nodes. By leveraging matrix perturbation theory and spectral analysis, our method prioritizes nodes according to their impact on the network’s Laplacian eigenvalues, thereby enhancing both synchronizability and convergence rates. Extensive experiments across various synthetic and real-world networks demonstrate that the proposed strategy consistently outperforms traditional centrality-based methods and closely approximates the performance of the computationally expensive brute-force greedy approach. Notably, in some generated networks and certain real-world networks, our method even surpasses the brute-force greedy strategy, highlighting its practical effectiveness and scalability in complex systems. 

Furthermore, our robustness analysis confirms that the proposed approach maintains high synchronizability even under node failure scenarios. Additionally, our theoretical findings establish a deeper connection between synchronizability and convergence rate, contributing to a broader understanding of synchronization dynamics in complex networks.

Looking ahead, several avenues for future research remain. First, extending the proposed method to directed and weighted networks is essential, as such structures are prevalent in real-world applications. Second, further investigation is needed to uncover the deeper relation between network sparsity and specific pinning strategies. Understanding these underlying mechanisms could lead to refined algorithmic enhancements and broader applicability in real-world synchronization-critical systems. Third, a deeper investigation into the relationships among synchronizability, spectral properties, and topological features is warranted, which may yield further insight into the design of more efficient pinning control strategies.

\section*{CRediT authorship contribution statement}
\textbf{Ziang Mao:} Conceptualization, Methodology, Formal analysis, Investigation, Writing - original draft, Writing - review \& editing, Visualization.
\textbf{Tianlong Fan:} Conceptualization,  Formal analysis, Writing - original draft, Writing - review \& editing, Supervision.
\textbf{Linyuan {Lü}:} Conceptualization, Writing - review \& editing, Supervision, Funding acquisition.

\section*{Data\&code availability}
The code, models and data presented in this study are available on request from the corresponding author.

\section*{Competing interests}
The authors declare no competing interests.

\section*{Acknowledgments}
This study was supported by the National Natural Science Foundation of China (Grant No. T2293771), the China Postdoctoral Science Foundation (Grant No. 2024M763131), the STI 2030—Major Projects (Grant No. 2022ZD0211400), the Postdoctoral Fellowship Program of CPSF (Grant No. GZC20241653) and the New Cornerstone Science Foundation through the XPLORER PRIZE.



 \bibliographystyle{elsarticle-num} 
 \bibliography{cas-refs}

\begin{thebibliography}{10}
\expandafter\ifx\csname url\endcsname\relax
  \def\url#1{\texttt{#1}}\fi
\expandafter\ifx\csname urlprefix\endcsname\relax\def\urlprefix{URL }\fi
\expandafter\ifx\csname href\endcsname\relax
  \def\href#1#2{#2} \def\path#1{#1}\fi

\bibitem{chen2015saturated}
C.~Chen, Z.~Liu, Y.~Zhang, C.~P. Chen, S.~Xie, Saturated nussbaum function based approach for robotic systems with unknown actuator dynamics, IEEE Transactions on Cybernetics 46~(10) (2015) 2311--2322.

\bibitem{dorfler2013synchronization}
F.~D{\"o}rfler, M.~Chertkov, F.~Bullo, Synchronization in complex oscillator networks and smart grids, Proceedings of the National Academy of Sciences 110~(6) (2013) 2005--2010.

\bibitem{zanette2002dynamics}
D.~H. Zanette, Dynamics of rumor propagation on small-world networks, Physical Review E 65~(4) (2002) 041908.

\bibitem{sorrentino2007controllability}
F.~Sorrentino, M.~Di~Bernardo, F.~Garofalo, G.~Chen, Controllability of complex networks via pinning, Physical Review E—Statistical, Nonlinear, and Soft Matter Physics 75~(4) (2007) 046103.

\bibitem{yu2013step}
W.~Yu, J.~L{\"u}, X.~Yu, G.~Chen, A step forward to pinning control of complex networks: Finding an optimal vertex to control, in: 2013 9th Asian Control Conference (ASCC), IEEE, 2013, pp. 1--6.

\bibitem{yu2013synchronization}
W.~Yu, G.~Chen, J.~Lu, J.~Kurths, Synchronization via pinning control on general complex networks, SIAM Journal on Control and Optimization 51~(2) (2013) 1395--1416.

\bibitem{brandes2001faster}
U.~Brandes, A faster algorithm for betweenness centrality, Journal of Mathematical Sociology 25~(2) (2001) 163--177.

\bibitem{wang2010control}
X.~Wang, X.~Li, J.~Lu, Control and flocking of networked systems via pinning, IEEE Circuits and Systems Magazine 10~(3) (2010) 83--91.

\bibitem{fan2021characterizing}
T.~Fan, L.~L{\"u}, D.~Shi, T.~Zhou, Characterizing cycle structure in complex networks, Communications Physics 4~(1) (2021) 272.

\bibitem{qiu2021identifying}
Z.~Qiu, T.~Fan, M.~Li, L.~L{\"u}, Identifying vital nodes by achlioptas process, New Journal of Physics 23~(3) (2021) 033036.

\bibitem{stewart1990stochastic}
G.~W. Stewart, Stochastic perturbation theory, SIAM Review 32~(4) (1990) 579--610.

\bibitem{miekkala1993graph}
U.~Miekkala, Graph properties for splitting with grounded laplacian matrices, BIT Numerical Mathematics 33~(3) (1993) 485--495.

\bibitem{pirani2014spectral}
M.~Pirani, S.~Sundaram, Spectral properties of the grounded laplacian matrix with applications to consensus in the presence of stubborn agents, in: 2014 American Control Conference, IEEE, 2014, pp. 2160--2165.

\bibitem{li2004pinning}
X.~Li, X.~Wang, G.~Chen, Pinning a complex dynamical network to its equilibrium, IEEE Transactions on Circuits and Systems I: Regular Papers 51~(10) (2004) 2074--2087.

\bibitem{yu2009pinning}
W.~Yu, G.~Chen, J.~L{\"u}, On pinning synchronization of complex dynamical networks, Automatica 45~(2) (2009) 429--435.

\bibitem{almendral2007dynamical}
J.~A. Almendral, A.~D{\'\i}az-Guilera, Dynamical and spectral properties of complex networks, New Journal of Physics 9~(6) (2007) 187.

\bibitem{amani2018controllability}
A.~M. Amani, M.~Jalili, X.~Yu, L.~Stone, Controllability of complex networks: Choosing the best driver set, Physical Review E 98~(3) (2018) 030302.

\bibitem{jalili2015optimal}
M.~Jalili, O.~Askari~Sichani, X.~Yu, Optimal pinning controllability of complex networks: Dependence on network structure, Physical Review E 91~(1) (2015) 012803.

\bibitem{jafarizadeh2021optimal}
S.~Jafarizadeh, D.~Veitch, F.~Tofigh, J.~Lipman, M.~Abolhasan, Optimal synchronizability in networks of coupled systems: Topological view, IEEE Transactions on Network Science and Engineering 8~(2) (2021) 1517--1530.

\bibitem{jafarizadeh2022pinning}
S.~Jafarizadeh, Pinning control of dynamical networks with optimal convergence rate, IEEE Transactions on Systems, Man, and Cybernetics: Systems 52~(11) (2022) 7160--7172.

\bibitem{cheng2018selecting}
Z.~Cheng, Y.~Xin, J.~Cao, X.~Yu, G.~Lu, Selecting pinning nodes to control complex networked systems, Science China Technological Sciences 61 (2018) 1537--1545.

\bibitem{liu2015synchronization}
X.~Liu, T.~Chen, Synchronization of complex networks via aperiodically intermittent pinning control, IEEE Transactions on Automatic Control 60~(12) (2015) 3316--3321.

\bibitem{xu2017finite}
C.~Xu, X.~Yang, J.~Lu, J.~Feng, F.~E. Alsaadi, T.~Hayat, Finite-time synchronization of networks via quantized intermittent pinning control, IEEE Transactions on Cybernetics 48~(10) (2017) 3021--3027.

\bibitem{ding2020synchronization}
S.~Ding, Z.~Wang, Synchronization of coupled neural networks via an event-dependent intermittent pinning control, IEEE Transactions on Systems, Man, and Cybernetics: Systems 52~(3) (2020) 1928--1934.

\bibitem{adaldo2015event}
A.~Adaldo, F.~Alderisio, D.~Liuzza, G.~Shi, D.~V. Dimarogonas, M.~Di~Bernardo, K.~H. Johansson, Event-triggered pinning control of switching networks, IEEE Transactions on Control of Network Systems 2~(2) (2015) 204--213.

\bibitem{li2018event}
B.~Li, Z.~Wang, L.~Ma, An event-triggered pinning control approach to synchronization of discrete-time stochastic complex dynamical networks, IEEE Transactions on Neural Networks and Learning Systems 29~(12) (2018) 5812--5822.

\bibitem{xing2020robust}
W.~Xing, P.~Shi, R.~K. Agarwal, L.~Li, Robust ${H}_{\ensuremath{\infty}}$ pinning synchronization for complex networks with event-triggered communication scheme, IEEE Transactions on Circuits and Systems I: Regular Papers 67~(12) (2020) 5233--5245.

\bibitem{li2016fuzzy}
X.-J. Li, G.-H. Yang, Fuzzy approximation-based global pinning synchronization control of uncertain complex dynamical networks, IEEE Transactions on Cybernetics 47~(4) (2016) 873--883.

\bibitem{nguyen2019fuzzy}
A.-T. Nguyen, T.~Taniguchi, L.~Eciolaza, V.~Campos, R.~Palhares, M.~Sugeno, Fuzzy control systems: Past, present and future, IEEE Computational Intelligence Magazine 14~(1) (2019) 56--68.

\bibitem{pecora1998master}
L.~M. Pecora, T.~L. Carroll, Master stability functions for synchronized coupled systems, Physical Review Letters 80~(10) (1998) 2109.

\bibitem{arenas2008synchronization}
A.~Arenas, A.~D{\'\i}az-Guilera, J.~Kurths, Y.~Moreno, C.~Zhou, Synchronization in complex networks, Physics Reports 469~(3) (2008) 93--153.

\bibitem{liu2018optimizing}
H.~Liu, X.~Xu, J.-A. Lu, G.~Chen, Z.~Zeng, Optimizing pinning control of complex dynamical networks based on spectral properties of grounded laplacian matrices, IEEE Transactions on Systems, Man, and Cybernetics: Systems 51~(2) (2018) 786--796.

\bibitem{donetti2005entangled}
L.~Donetti, P.~I. Hurtado, M.~A. Munoz, Entangled networks, synchronization, and optimal network topology, Physical Review Letters 95~(18) (2005) 188701.

\bibitem{tang2014synchronization}
Y.~Tang, F.~Qian, H.~Gao, J.~Kurths, Synchronization in complex networks and its application--a survey of recent advances and challenges, Annual Reviews in Control 38~(2) (2014) 184--198.

\bibitem{yan2009synchronization}
G.~Yan, G.~Chen, J.~L{\"u}, Z.-Q. Fu, Synchronization performance of complex oscillator networks, Physical Review E—Statistical, Nonlinear, and Soft Matter Physics 80~(5) (2009) 056116.

\bibitem{albert2002statistical}
R.~Albert, A.-L. Barab{\'a}si, Statistical mechanics of complex networks, Reviews of Modern Physics 74~(1) (2002) 47.

\bibitem{watts1998collective}
D.~J. Watts, S.~H. Strogatz, Collective dynamics of ‘small-world’networks, Nature 393~(6684) (1998) 440--442.

\bibitem{rossi2015network}
R.~Rossi, N.~Ahmed, The network data repository with interactive graph analytics and visualization, in: Proceedings of the AAAI conference on artificial intelligence, Vol.~29, 2015.

\bibitem{duch2005community}
J.~Duch, A.~Arenas, Community identification using extremal optimization phys, Rev. E 72 (2005) 027104.

\bibitem{ahmed2010time}
N.~Ahmed, F.~Berchmans, J.~Neville, R.~Kompella, Time-based sampling of social network activity graphs, in: SIGKDD MLG, 2010, pp. 1--9.

\bibitem{infect}
{SocioPatterns}, {SocioPatterns Datasets}, \url{http://www.sociopatterns.org/datasets/}, accessed: 2025-03-11.

\bibitem{traud:2011fs}
A.~L. Traud, E.~D. Kelsic, P.~J. Mucha, M.~A. Porter, Comparing community structure to characteristics in online collegiate social networks, SIAM Rev. 53~(3) (2011) 526--543.

\bibitem{traud2012social}
A.~L. Traud, P.~J. Mucha, M.~A. Porter, Social structure of {F}acebook networks, Phys. A 391~(16) (2012) 4165--4180.

\bibitem{zhou2008pinning}
J.~Zhou, J.-a. Lu, J.~L{\"u}, Pinning adaptive synchronization of a general complex dynamical network, Automatica 44~(4) (2008) 996--1003.

\bibitem{li2020secure}
Y.~Li, D.~Shi, T.~Chen, Secure analysis of dynamic networks under pinning attacks against synchronization, Automatica 111 (2020) 108576.

\bibitem{lou2021knowledge}
Y.~Lou, Y.~He, L.~Wang, K.~F. Tsang, G.~Chen, Knowledge-based prediction of network controllability robustness, IEEE Transactions on Neural Networks and Learning Systems 33~(10) (2021) 5739--5750.

\end{thebibliography}





\end{document}